\documentclass[a4paper,10pt, onecolumn]{mn}

\usepackage[utf8x]{inputenc}
\usepackage{amsmath}
\usepackage{amssymb}
\usepackage{color}
\usepackage{subfig}
\usepackage{graphicx}
\graphicspath{{figures/}}
\usepackage{dcolumn}
\usepackage{bm}
\usepackage{mathrsfs}
\usepackage{natbib}
\usepackage{textcomp}
\usepackage{tikz}
\usetikzlibrary{arrows,shapes,trees,snakes}
\newcommand{\boundellipse}[3]
{(#1) ellipse (#2 and #3)
}

\newcommand{\be}{\begin{equation}}
\newcommand{\ee}{\end{equation}}
\newcommand{\ba}{\begin{align}}
\newcommand{\ea}{\end{align}}

\newcommand{\expo}{{\rm exp}}

\newcommand{\gae}{\lower 2pt \hbox{$\, \buildrel {\scriptstyle >}\over {\scriptstyle \sim}\,$}}
\newcommand{\lae}{\lower 2pt \hbox{$\, \buildrel {\scriptstyle <}\over {\scriptstyle \sim}\,$}}
\newcommand{\aprop}{\lower 2pt \hbox{$\, \buildrel {\scriptstyle \propto}\over 
   {\scriptstyle \sim}\,$}}

\begin{document}

\title[IC drag on Highly Magnetized GRB jets]{Inverse-Compton drag on a Highly Magnetized GRB jet in Stellar Envelope}

\author[Ceccobello and Kumar]
{Chiara Ceccobello$^{1,2}$, Pawan\ Kumar$^{3}$
\thanks{E-mail:  c.ceccobello@uva.nl; pk@astro.as.utexas.edu} \\
$^{1}$``Anton Pannekoek'' Instituut voor Sterrekunde, Universiteit van Amsterdam, Science Park 904, 1098 XH Amsterdam, The Netherlands\\
$^{2}$Dipartimento di Fisica, Universit\`a di Ferrara, via Saragat 1, 44122, Ferrara, Italy\\
$^{3}$\textit{Department of Astronomy, University of Texas at Austin, 
 TX 78712, USA }}

\date{Accepted 2015 March 2. Received 2014 July 16}
\maketitle

\begin{abstract}
The collimation and evolution of relativistic outflows in $\gamma$-ray bursts 
(GRBs) are determined by their interaction with the stellar envelope through 
which they travel before reaching the much larger distance where the energy is 
dissipated and $\gamma$-rays are produced. We consider the case of a Poynting 
flux dominated relativistic outflow and show that it suffers strong 
inverse-Compton (IC) scattering drag near the stellar surface and the jet 
is slowed down to sub-relativistic speed if its initial 
magnetization parameter ($\sigma_0$) is larger than about 10$^5$. 
If the temperature of the cocoon surrounding the jet were to be larger 
than about 10 keV, then an optically thick layer of electrons and 
positrons forms at the interface of the cocoon and the jet, and one might
expect this pair screen to protect the interior of the jet from IC drag.
However, the pair screen turns out to be ephemeral, and instead of
shielding the jet it speeds up the IC drag on it. 
Although a high $\sigma_0$ jet might not survive its passage through 
the star, a fraction of its energy is converted to 1-100 MeV radiation 
that escapes the star and appears as a bright flash lasting for about 10 s.
\end{abstract}


\section{Introduction}

Long Gamma Ray Bursts (long GRBs) are explosions resulting from the core
collapse of massive stars at the end of their nuclear burning life cycle. 
The amount of energy produced in these explosions is estimated to be 
$\sim10^{48}$--$10^{52}$ erg, e.g. \cite{Sari:1999,Frail:2001,PanaitescuKumar:2001,Berger:2003,Curran:2008,Liang:2008,Racusin:2009,Cenko:2010}.
The collapse of the core of a GRB progenitor produces either a black hole 
or a neutron star, and in either case the central compact object is believed to 
be rapidly rotating (for a review \cite{Piran:1999,Meszaros:2006,Woosley:2006,Gehrels:2009}). 
As in other astrophysical sources such as active galactic nuclei, 
micro-quasars, pulsars and SGRs (soft-gamma-ray repeaters) a rapidly 
rotating black hole, or a magnetar, is expected to produce a relativistic 
bipolar jet which then interact with the ambient medium (e.g. 
\cite{Bucciantini:2008,Bucciantini:2009},\cite{Falcke:2004},
\cite{Markoff:2005,Markoff:2010}, \cite{Narayan:2008}, \cite{YuanNarayan}).

In the scenario where the central engine of a GRB is a rapidly rotating 
magnetar, a Poynting-flux dominated jet is generated by the strong magnetic 
field with an initial jet magnetization parameter\footnote{Magnetization
parameter is defined as the ratio of Poynting flux and particle kinetic energy
flux carried by the jet.}, $\sigma_0$, of order 
$\sim10^3$ (\cite{Thompson:2004,Metzger:2007}). The magnetization parameter 
increases as the neutrino driven baryonic mass loss rate at the surface of 
the neutron star decreases on de-leptonization timescale of about half a minute 
\citep{Metzger:2011}. In fact, the increase to the magnetization parameter 
can be rather dramatic with $\sigma_0\sim 10^9$ as the neutrino luminosity 
winds down \citep{Metzger:2011}. These authors associate the transition
to high $\sigma_0$ with the end of the prompt $\gamma$-ray phase and the 
steep decline of X-ray afterglow that is seen for a large fraction of 
bursts detected by the Swift satellite (\cite{Tagliaferri:2005,OBrien:2006,Willingale:2010}). The reason for this association, according to \cite{Metzger:2011},
is that the acceleration and dissipation are very inefficient processes for 
high $\sigma_0$ jets. It should be noted that highly magnetized jets are
not limited to the magnetar model, but could also be produced when the GRB 
central engine is an accreting black hole as the mechanism for launching of
the jet might be the Blandford-Znajek process \citep{BlandfordZnajek:1977}.
In this paper we address the question regarding the survival of a highly
magnetized jet ($\sigma_0\gae10^4$) as it propagates through the GRB 
progenitor star and is exposed an intense radiation field that can penetrate
all the way to jet axis and can drag it down.
IC drag has been considered by a number of people for AGN jets e.g. \cite{Phinney:1987,MeliaKonigl:1989,Sikora:1996, Ghisellini:2005, GhiselliniTavecchio:2010}.
However, the GRB jets are different in that they are highly opaque throughout much of the region where they undergo acceleration (as opposed to AGN jets which are transparent throughout their entire length), and thus the IC drag on GRB
jets due to an external field can be treated independently of the acceleration mechanism.\\
A jet propagating through the GRB progenitor star envelope creates a hot cocoon
within which the jet is enclosed by the time it reaches the stellar surface.
We describe the interaction of such highly magnetized jets with photons from the
hot cocoon and how that affects the jet propagation through the envelope of 
the progenitor star.
In particular, we analyze the role of the inverse Compton drag in braking 
the relativistic outflow. 

\noindent 
The organization of this paper is as follows.
In the next section we calculate how far inside the jet can X-ray photons
from the cocoon penetrate the jet, and at the distance from the center
when the jet becomes transparent. 
\S3 provides an estimate for IC drag on a high magnetization 
parameter jet due to scattering of X-ray photons from the cocoon surrounding 
it, but ignoring the creation of $e^\pm$ pairs and possible shielding of the 
jet provided by these particles. In \S4 we consider whether the core of the 
jet could be shielded by electron-positron pairs produced either at the base 
of the jet or at the interface of the jet and cocoon by collision of X-ray
photons with $\gamma$-ray photons that are arise when X-ray photons are 
inverse-Compton (IC) scattered by electrons in the jet. 

\section{Photospheric radius for Poynting flux dominated jets}

Let us consider a relativistic Poynting-dominated jet composed by a mixture of baryons, leptons and photons. As seen in the frame comoving with the flow, baryons and leptons are thermally distributed and have the same number density.

The total isotropic equivalent luminosity of a Poynting jet, which has 
significant thermal energy, and where mass flux is dominated by baryons, can 
be written as
\be
  L = {\mathrm \pi} R^2\theta_j^2(R)\left[m_p n_p'  v c^2 \Gamma^2 + {4\over 3}u_\gamma' 
   \Gamma^2 v + {B'^2\Gamma^2 v\over 4{\mathrm \pi}} \right],
\ee
where $\Gamma$ and $v$ are jet Lorentz factor and speed, $n_p'$ and $B'$ are 
jet comoving frame proton density and magnetic field strength, $u'_\gamma$
is the energy density in photons in jet comoving frame, and $\theta_j(R)$
is the jet half-opening angle when it is at radius $R$.\\
Using the mass conservation equation and the definition of the magnetization parameter, which are the following
\begin{align}
 \dot M_\pm &= {\mathrm \pi} \theta_j^2  R^2 m_p n_p' \Gamma v  \label{m_dot}\\
 \sigma &= \frac{B'^2}{4{\mathrm \pi} m_p n_p' c^2},   \label{mag_para}
\end{align}
we obtain
\begin{equation}
 L = \dot M_\pm c^2\Gamma\left(1+\xi+\sigma\right),
   \label{lee_mag}
\end{equation}
where
\be
  \xi \equiv {4 u_\gamma'\over 3 m_p n_p' c^2}
\ee
is the ratio of thermal energy and baryon rest-mass energy densities.
From equations (\ref{m_dot})--(\ref{lee_mag}) we calculate proton number 
density in the comoving frame
\be
n_p'(R)\approx \frac{L}{\Gamma_0(1 + \xi_0 + \sigma_0) c^2}\frac{1}{{\mathrm \pi}
 \theta_j^2 R^2 m_p c \Gamma} \approx \frac{L}{{\mathrm \pi}\theta_j^2 R^2 m_p c^3 
  (\sigma_0 + \xi_0)\Gamma}, 
   \label{np-prime}
\ee
where $\Gamma_0$, $\xi_0$ and $\sigma_0$ are values at the base of the jet
at radius $R_0$, and the last part of the above equation is obtained by
assuming that $\Gamma_0\sim1$ and $\sigma_0 + \xi_0\gg 1$.

While the jet is inside the star it is collimated by the pressure of the
cocoon and the ambient stellar medium, and we take its Lorentz factor to 
increase with radius as 
\be
  \Gamma \sim \left[ {r\over R_0}\right]^\alpha,
  \label{gam_r}
\ee
as long as $\Gamma < (\xi_0 + \sigma_0)$. The index $\alpha$ in the 
above equation can be shown to be related to the pressure ($p$)
stratification of GRB progenitor star; for $p\propto r^{-a}$, $\alpha=a/4$
as long as $a \le 2$ \citep{KumarZhang:2014}. The pressure in the He-envelope 
of the GRB progenitor star declines as $\sim r^{-2}$, and therefore we
expect $\alpha\sim 0.5$ for a Poynting jet. The transverse size of the 
jet increases with radius in the same way as $\Gamma$  \citep{KumarZhang:2014}, 
i.e. 
\be
  R_{j,\perp}(r) \sim R_0 \left[ {r\over R_0}\right]^\alpha,
  \label{R_perp}
\ee
and therefore the jet angle is given by
\be
   \theta(r) \sim \left[ {R_0\over r}\right]^{1-\alpha}.
   \label{theta_r}
\ee

The photospheric radius where the Thompson optical depth of the jet 
in the jet-longitudinal direction is unity, is determined from
\be
\tau_T \approx \int \frac{dr}{2\Gamma^2} \sigma_T n_p'(r)\Gamma 
    \approx {\sigma_T L\over 2{\mathrm \pi} (4\alpha-1) R_0 m_p c^3 (\sigma_0 + \xi_0)}
    \left[{R_0\over R}\right]^{4\alpha-1},
   \label{tauT}
\ee
where we made use of equation (\ref{np-prime}) for electron number density 
(which is assumed to be equal to proton number density), and equations 
(\ref{gam_r}) \& (\ref{theta_r}) to substitute for $\Gamma$ and $\theta_j$
respectively. Thus the photospheric radius is
\begin{equation}
  R_{ph} = R_0 \left[ \frac{\sigma_T L}{2{\mathrm \pi} (4\alpha-1) m_p c^3 
   (\sigma_0 + \xi_0)R_0}\right]^{{1\over 4\alpha-1}} \sim \left(2.4\times 
   10^{9} \rm{cm}\right) L_{50}\, \sigma_{0,6}^{-1}\, (\Gamma_2\theta_{j,-1})^{-2}.
 \label{ph_r_mag}
\end{equation}
So in case of initially large magnetization parameter $(\gtrsim 10^4)$,  
photons can escape from the jet well before the jet reaches the stellar
surface. We note that the maximum Lorentz factor of a thermal fireball with 
$\sigma_0=0$ is obtained when the jet acceleration continues out to the
photospheric radius $R_{ph}$ and is given by 
$\Gamma_{max}\sim [\sigma_T L/(2{\mathrm \pi} (4\alpha-1)m_p c^3 R_0)]^{\alpha/(5\alpha-1)}$ 
as long as $R_{ph} < R_*$; for $\alpha=1$, 
$\Gamma_{max}\sim 5.4\times10^2L_{51}^{1/4}R_{0,7}^{-1/4}$.

Photons from the cocoon surrounding the jet cannot penetrate very far 
inside the jet at radius $R_{ph}$ because of the much larger optical depth 
in the transverse direction. To estimate the inverse-Compton (IC) drag
of the jet due to scattering of X-ray photons from the cocoon by electrons
in the jet we first determine the radius where the jet becomes transparent
in the transverse direction.

\subsection{Jet transparency radius in the transverse direction}

The optical depth of the jet for photons of frequency $\nu_c$ traveling 
in a direction perpendicular to the jet axis, including Klein-Nishina 
correction to the scattering cross-section, is 
\begin{equation}
 \tau_{\bot}(r) \approx {\sigma_T n_p'(r)  \Gamma\theta_j r \over 1 + h\nu_c
   \Gamma/(m_e c^2)} \left[ 1 + 2\ln\left(1 + {2 h\nu_c\Gamma\over m_e c^2}
   \right)\right] \approx \frac{\sigma_T L}{{\mathrm \pi} \theta_j m_p c^3 
   (\sigma_0 + \xi_0) r [1 + h\nu_c\Gamma/(m_e c^2)]},
    \label{tau_T}
\end{equation}
where we used equation (\ref{np-prime}) for $n_p'$.
The radius where the jet becomes transparent to photons moving in 
the transverse direction is
\begin{equation}
  R_{ph,\bot}\approx\left(5\times10^{12}\rm{cm}\right) {L_{50} 
  \sigma_{0,6}^{-1} \over \theta_{j,-1} [1 + h\nu_c\Gamma/(m_e c^2)]}
   \label{r_ph}.
\end{equation}

A more general situation is where photons are traveling at an angle 
$\theta_\gamma$ with respect to the jet axis. We calculate the optical depth for these 
photons to travel from the interface of jet and the cocoon to the jet axis.

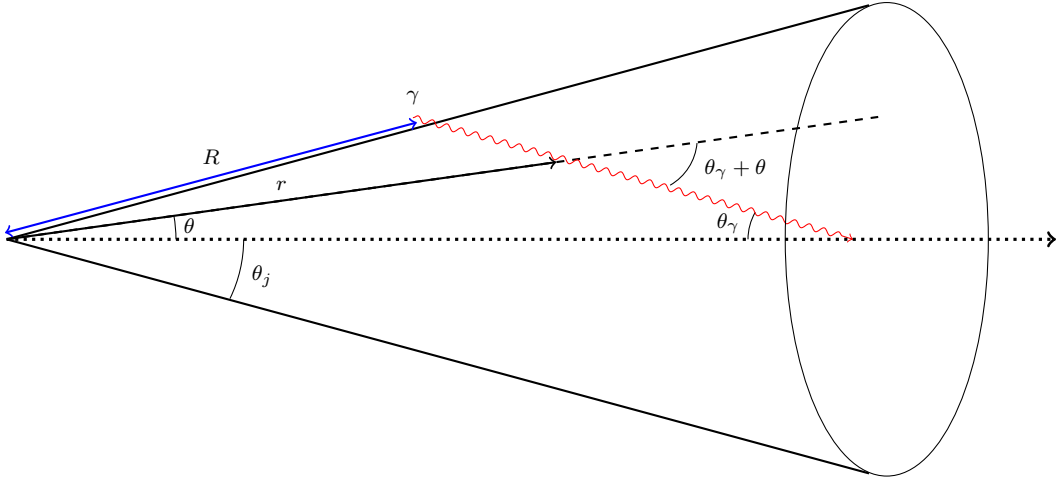
\begin{figure}
\centering
\begin{tikzpicture}[scale=0.9]
\draw[black,solid, thick,rotate=15.2] (0,0) -- (13.2,0);
\draw[blue,solid, thick,rotate=15,<-> ] (0,0.1) -- node[above=2pt,color=black] {$R$} (6.3,0.1);
\draw[black, dashed, thick,rotate=8] (0,0) -- (13,0);
\draw[black, solid, thick,rotate=8,->] (0,0) -- node[above=1pt,color=black] {$r$} (8.2,0);
\draw (10.95,0)  node[above=6pt,left,color=black] {$\theta_\gamma$} arc[radius = .8cm, start angle= 180, end angle=150];
\draw (9.8,.8)  node[above=6pt,right=10pt,color=black] {$\theta_\gamma+\theta$} arc[radius = .8cm, start angle= 300, end angle=355];
\draw[->,red,snake=snake,segment amplitude=.4mm,segment length=2mm,line after snake=1mm] (6,1.8) node[above=2pt,color=black] {$\gamma$} -- (12.5,0);
\draw[black, dotted, very thick,->] (0,0) -- (15.5,0);
\draw[black,solid, thick,rotate=-15.2] (0,0) -- (13.2,0);
\draw \boundellipse{13,0}{-1.5}{3.5};
\draw (2.5,0) node[above=5pt, right,color=black] {$\theta$} arc[radius = 2cm, start angle= 0, end angle=10];
\draw (3.5,0) node[below=14pt,right,color=black] {$\theta_j$} arc[radius = 2cm, start angle= 0, end angle=-26];
\end{tikzpicture}
\caption{Schematic sketch of a photon trajectory (red line) from the hot cocoon 
passing through the jet axis (dotted line). The angle between the photon initial
 direction of motion and the jet axis is labeled as $\theta_\gamma$ and the jet 
opening angle is $\theta_j$. $R$ is the radial distance between the center
of explosion and the injection point where the cocoon photon enters the jet 
and $r$ is the radial coordinate. Although the jet is shown to be conical in 
this representation, its transverse radius increases more slowly than $r$ 
while it is confined by the pressure of the star/cocoon. }
\label{FIG:cc-photon-trajectory}
\end{figure}
Consider a photon traveling at an angle $\theta_\gamma$ with respect to jet axis 
and electrons moving in the radial direction (see Fig. 
\ref{FIG:cc-photon-trajectory}). The optical depth for a photon
of frequency $\nu_c$ to scatter off electrons along its trajectory 
starting from cocoon-jet interface to the jet axis is
\begin{equation}
   \tau(\theta_\gamma) \approx \int_0^{\theta_j} d\theta\,r\, {\sigma_T n_p(r) 
   [1 - v\cos(\theta+\theta_\gamma)/c] \over 
    [1 + h\nu_c'/m_e c^2] \sin(\theta+\theta_\gamma)},
   \label{tau2a}
\end{equation}
where $n_p = \Gamma n_p'$ is electron density in star rest-frame, 
\be 
  \nu_c'(\theta+\theta_\gamma) = \nu_c \Gamma\left[ 1 - v\cos(\theta +
   \theta_\gamma)/c \right],
\ee
is photon frequency in electron rest frame, and we have assumed that there 
is one electron per proton in the jet; Klein-Nishina correction to 
Thompson cross-section is included in eq. (\ref{tau2a}). The photon trajectory 
in the polar coordinate is described by (see Fig. 
\ref{FIG:cc-photon-trajectory})
\begin{equation}
   r\sin(\theta+\theta_\gamma) = R\sin(\theta_j + \theta_\gamma).
   \label{photon-trajectory}
\end{equation} 
Using 
\begin{equation}
   n_p(r) = {L\over {\mathrm \pi} \theta_j^2 r^2 m_p c^3 (\sigma_0+\xi_0)}=(7\times10^{17}
   {\rm cm}^{-3}) {L_{50} \over \theta_{j,-1}^2 r_{10}^{2} (\sigma_{0,6}
   + \xi_{0,6})},
   \label{np}
\end{equation}
we find
\begin{equation}
   \tau = {\sigma_T L\over {\mathrm \pi}\theta_j^2 m_p c^3 (\sigma_0+\xi_0) }
    \int_0^{\theta_j} \, d\theta 
    { \left[ {1 - v \cos(\theta+\theta_\gamma)/c}\right] \over r 
    [1 + h\nu_c'(\theta+\theta_\gamma)/m_e c^2] \sin(\theta+\theta_\gamma) }.
   \label{tau1}
\end{equation}
With use of equation (\ref{photon-trajectory}) for $r$ this reduces to
\begin{equation}
   \tau(r) \approx {\sigma_T L \left[ 1 + h \nu_c'(\theta_\gamma)/m_e c^2
   \right]^{-1} \over {\mathrm \pi} \theta_j^2 m_p c^3 (\sigma_0+\xi_0) r\sin(\theta_j + 
  \theta_\gamma) } \left[ \theta_j - v \left\{\sin(\theta_j+\theta_\gamma) 
  - \sin\theta_\gamma \right\}/c\right]. 
   \label{tau2}
\end{equation}

The GRB jet Lorentz factor near the surface of the progenitor star is 
expected to be much larger than $\theta_j^{-1}$. In that case for 
$\theta_\gamma \ll \theta_j$ equation (\ref{tau2}) reduces to
\begin{equation}
   \tau \approx 1.5 L_{50} (\sigma_{0,6}+\xi_{0,6})^{-1}
   r_{11}^{-1} \left[ 1 + h \nu_c'(\theta_\gamma)/m_e c^2 \right]^{-1},
   \label{optical-depth}
\end{equation}
and for $\theta_\gamma \gg \theta_j$ the expression for optical depth
reduces to equation (\ref{tau_T}). The jet optical depth in the 
transverse direction at a given radius for these two limiting cases differs
 by a factor $\sim 10$.

\section{Inverse Compton drag for Poynting flux dominated jets}

We calculate the inverse-Compton loss for an electron when it is exposed 
to a beam of photons moving at an angle $\theta_\gamma$ with respect to 
electron velocity. The electron moves with the jet and therefore its Lorentz 
factor and velocity are $\Gamma$ and $v$ respectively.  Let us consider
the specific intensity of the photon beam from the cocoon in the rest frame 
of the star to be $I_\nu(\theta_\gamma)$. The
transformations of photon frequency, specific intensity, and angle from 
star-rest frame to jet comoving frame are given by
\begin{equation}
   \nu' = \nu \Gamma(1 - v\cos\theta_\gamma/c), \quad\quad I'_{\nu'}(\theta') = 
    I_\nu (\nu'/\nu)^3, \quad\quad \sin\theta' = (\nu/\nu')\sin\theta_\gamma.
\end{equation}
The rate of loss of energy for an electron due to inverse-Compton scatterings
is proportional to photon energy flux in its comoving frame. Using the
above transformations the comoving flux can be shown to be proportional to
$\theta_\gamma^2 \Gamma^2(1 - v\cos\theta_\gamma/c)^2\propto \theta_\gamma^6
\Gamma^2$ for $\Gamma\theta_\gamma > 1$. Thus, the IC drag on electrons 
increases very rapidly with increasing $\theta_\gamma$, and that nearly 
compensates for the smaller flux at the jet axis due to larger optical 
depth for larger $\theta_\gamma$. So from here on we specialize to 
$\theta_\gamma\sim 1$.
Consider a photon of frequency $\nu_c$ scattered by an electron of Lorentz factor
$\Gamma$. 
In the comoving frame, the scattered photon energy is
\begin{equation}
    h\nu_{ic}' = {h\nu_c' \over 1+{h\nu_c' \over m_e c^2}(1-\cos\theta_\gamma)}.
\end{equation}
The average energy of the scattered photon in the Klein-Nishina limit can be calculated as
\begin{equation}
    \langle h\nu_{ic}\rangle  = {\int d\Omega_\gamma' {d\sigma_{kn}\over d\Omega_\gamma'} h\nu_{ic} \over \int d\Omega_\gamma' {d\sigma_{kn}\over d\Omega_\gamma'}}
\end{equation}
where the differential Klein-Nishina cross section is taken from \citep{BlumenthalGould:1970}. When $2\lesssim h \nu_c \Gamma/(m_e c^2) \lesssim \Gamma^2$, the average energy of the scattered photon can be shown to be
\begin{equation}
    \langle h\nu_{ic}\rangle \approx {2 m_e c^2\Gamma \over 1 + 2 \ln[1 + 2 h\nu_c\Gamma/(m_e c^2)]} \left[ {3 m_e c^2\over 2 h\nu_c\Gamma} + \left( 1 - {m_e c^2\over h\nu_c\Gamma}\right)\ln\left( 1 + {h\nu_c\Gamma\over 2 m_e c^2}	    \right)\right].
\label{ic-nu1}
\end{equation}
The equation for IC drag is 
\be
   {d m_p c^2 \Gamma\sigma\over dt} \approx - {F_c(t)\over h\nu_c}\sigma_{kn}
     \,h \nu_{ic} \,\zeta_\pm,
    \label{ic-loss1}
\ee
where $\zeta_\pm$ is number of electrons and positrons per proton, $F_c$
is cocoon thermal flux given by equation (\ref{Fc}), and the electron-photon
scattering cross-section as a function of dimensionless photon energy in 
electron rest frame ($h\nu_c\Gamma/m_e c^2$) is
\be
  \sigma_{kn}(h\nu_c\Gamma/m_e c^2) \approx {3 \sigma_T \over 16 
   h\nu_c\Gamma/(m_e 
   c^2)} \left[ 1 + 2\ln\left( 1 + {2h\nu_c\Gamma\over m_e c^2}\right)\right].
   \label{sigma-kn}
\ee
In deriving equation (\ref{ic-loss1}) we assumed that electrons, protons 
and electro-magnetic fields are coupled and move together; charge particles
are coupled to the jet EM field via $\vec E\times\vec B$ drift force which
ensures that particles move with EM fields, and thus the total energy per 
proton in the jet is $\sim\sigma m_p c^2\Gamma$.

The IC cooling time in the stellar rest frame for an electron in the
jet, at radius $r$ and time $t_r$ after the formation of cocoon, follows 
from equations (\ref{ic-loss1}) and (\ref{sigma-kn}):
\be
  t_{ic}^{kn}\approx{8 m_p c^2\sigma_0(h\nu_c/m_e c^2)^2\over3\sigma_T\zeta_\pm 
   F_c(t_r) } \left[ {3 m_e c^2\over 2 h\nu_c\Gamma} + \left( 1 - {m_e c^2 
    \over h\nu_c \Gamma}\right)\ln\left(1 + {h\nu_c\Gamma\over 2 m_e c^2}
    \right)\right]^{-1}
   \label{t-ic1}
\ee
Substituting for $h\nu_c=3 k_B T_c$ (eq. \ref{Tc}) and $F_c$ (eq. \ref{Fc}) 
we obtain the IC cooling time in Klein-Nishina regime to be
\be
  t_{ic}^{kn} \sim (0.8{\rm s}) \exp(\tau_\perp) t_r^{1/2}
     \sigma_{0,6} \eta_{c,1}^{-1/2} \zeta^{-1}_\pm,
   \label{t-ic-kn}
\ee
where $\eta_c$ is the terminal Lorentz factor of the cocoon.
The IC cooling time when the scattering is in the Thompson regime,
i.e. $h \nu_c\Gamma \ll m_e c^2$, is given by
\be
   t_{ic}^{ts} \sim (5\times10^{-3}{\rm s}) \exp(\tau_\perp) t_r^{1/2}
     \Gamma_2^{-2} \sigma_{0,6} \eta_{c,1}^{-1/2} \zeta^{-1}_\pm.
   \label{t-ic-ts}
\ee

The dynamical time at the stellar surface is $R_*/c\sim 3$ s. The
IC cooling time is smaller than the dynamical time as long as
$\sigma_0\gae 10^5$.
Equation (\ref{t-ic1}) can be solved to find the radius, $R_{CC}$, where the IC 
drag timescale at the jet-axis ($t_{ic}$) is comparable to the dynamical 
timescale of the jet ($t_{dyn}\sim r/c$), namely 
\be
 \frac{8 m_p c^2 \sigma_0 t_r^{1/2} (h\nu_c/m_e c^2)^2}{3\sigma_T\sigma_B 
     T_c^4 t_{fs}^{1/2} e^{-\tau_\bot} \zeta_\pm} \approx \frac{R_{CC}}{c},
    \label{RCC1}
\ee
where $t_r$ is the time when IC cooling is considered --- it is in general
larger than the dynamical time since cocoon formation begins with the
launch of the relativistic jet and jet duration should exceed $R_*/c$
in order for the jet to break through the stellar surface  --- and 
$t_{fs}$ (given by eq. \ref{tfs}) in the mean time in between scatterings 
of a photon while inside the cocoon; the equation is valid for $h\nu_c\Gamma/
(m_e c^2) \gg 1$.

Making use of equation (\ref{tau_T}) for $\tau_\perp$ we can rewrite the 
above equation for $R_{CC}$ in a more explicit form:
\be
   \expo\left\{\frac{\sigma_T L}{{\mathrm \pi} \theta_j m_p c^3 (\sigma_0 + 
   \xi_0) R_{CC} [1 + h\nu_c\Gamma/(m_e c^2)]}\right\} = \frac{3\sigma_T
   \sigma_B T_c^4 t_{fs}^{1/2} \zeta_\pm R_{CC}}{8 m_p c^3 \sigma_0
   t_r^{1/2}(h\nu_c/m_e c^2)^2} 
    \label{rcc_eqn_explicit}.
\ee

Equation (\ref{rcc_eqn_explicit}) can be rewritten with the use of 
equation (\ref{gam_r}) for $\Gamma$ 
\be
e^{A/R_{CC}}=K\,R_{CC}^{\beta} \label{rcc_eqn_gen}
\ee
where 
\be
 A=\frac{\sigma_T L}{{\mathrm \pi} \theta_j m_p c^3 (\sigma_0+\xi_0) [1 + 
   h\nu_c\Gamma/(m_e c^2)]}, \quad\, K = \frac{3\sigma_T\sigma_B T_c^4 
   t_{fs}^{1/2} \zeta_\pm}{8 m_p c^3 \sigma_0 t_r^{1/2} (h\nu_c/m_e c^2)^2}. 
\ee
The equation for the Compton cooling radius $R_{CC}$ is a transcendental 
equation which can be solved perturbatively after we transform it into 
the following logarithmic transcendental equation
\be
\frac{A}{R_{CC}} = \log(K) + \log(R_{CC}) \label{rcc_log}
\ee
At the zero-\emph{th} order, $\log(R_{CC})$ term on the right side of the above 
equation is neglected and, the solution is
\be
R_{CC}^*=\frac{A}{\log(K)} \label{1st_ord_rcc},
\ee 
which, when substituted back into (\ref{rcc_log}), provides the first order solution for the Compton cooling radius
\be
  R_{CC} = \frac{A}{\log(K)+\left[\log(A)-\log(\log(AK))\right]}
  \label{rcc_an_sol}
\ee
The estimate for $R_{CC}$ using equation (\ref{rcc_an_sol}) and the exact solution of eq. \ref{rcc_eqn_gen} for parameters $L_{50}=1, \theta_j=0.1, 
T_{c,7}=1, \Gamma_2=1, \sigma_{0,6}=1$, and
$\zeta_\pm =1$ are $R_{CC} = 7.5\times 10^{11} 
{\rm cm}$ and $R_{CC}=7.0\times 10^{11} {\rm cm}$ respectively.
 

\section{Shielding jets from IC drag by creation of electrons and positrons}
\label{ep-p}

Thus far we have assumed that the jet energy is carried outward by 
magnetic fields, protons and electron-positron pairs. However, we have
not estimated the number of $e^\pm$ that might have been produced in
the hot plasma at the base of the jet, or pairs that might be created
in the collision of inverse-Compton scattered photons.
The presence of these pairs could shield the inner part of the jet from
IC drag due to thermal photons from the cocoon. In the next subsection
we take up the calculation of thermal $e^\pm$ that owe their existence
to the initial hot plasma at the base of the jet, and show that their
number density is too small far away from the jet launching site to 
be able to shield the jet. In section \ref{pair-create} we provide an estimate
of the density of pairs generated when thermal photons from the cocoon collide 
with photons that are IC scattered by $e^\pm$ in the jet. This process
is shown to be effective in shielding the jet for a while but eventually
pairs annihilate and pair screen disappears exposing the jet-core to 
severe IC drag (\S\ref{pair-create}).

\subsection{Thermal pairs and shielding of Poynting jets}

The number density of {\it thermal} pairs at any $r$ is given by the 
standard thermal distribution formula corresponding to the local temperature 
of the jet as long as the $e^\pm$ annihilation time is less than the dynamical 
time\footnote{A more precise statement is that number density of pairs at 
any given radius is given by the balance between creation and annihilation
rates. However, it can be shown that the assumption of thermal equilibrium
is approximately valid as long as annihilation time is short compared with
the expansion time.}. 
The radius where the two time scales become equal,
$R_{freeze}$, is the freeze-out radius for pairs. Beyond this radius
the total number of $e^\pm$ does not change barring the dissipation of
jet kinetic, or magnetic, energy and using that to create new pairs; 
non-thermal pair creation will be taken up in \S\ref{pair-create}. 
We calculate the number density of pairs at $R_{freeze}$ and show that 
to be much smaller than the density of protons. Therefore, thermal pairs
are unimportant for shielding the jet.

The temperature at the base of a Poynting jet is 
\be
 k_B T_0 \approx k_B \left[\frac{\xi_0 L^{iso}}{4{\mathrm \pi} R_0^2 \sigma_B 
     (\sigma_0+\xi_0)}\right]^{1/4} 
   \approx (41\,{\rm keV})\, (\xi_0 L_{52}^{iso}/\sigma_{0,6})^{1/4} 
   R_{0,7}^{-1/2} 
\ee
which is considerably smaller than the temperature for a thermal
fireball with $\sigma_0 < 1$; $\L^{iso} \equiv 4 L/\theta_j^2$ is
isotropic equivalent of jet luminosity ($\theta_j\sim 1$ at the
jet-base).

Considering the conservation of entropy in a shell of plasma as it moves 
to larger radius with the jet we find the decrease of comoving frame 
temperature with $r$
\be
 T'(r)
  \sim T_0 \left(\frac{R_0}{r}\right)^{2\alpha/3}\Gamma^{-1/3} \sim 
   T_0 \left(\frac{R_0}{r}\right)^{\alpha}.
    \label{t_erf_nolf}
\ee
where we have made use of equation (\ref{R_perp}) for the transverse size of 
the jet and equation (\ref{gam_r}) for $\Gamma$; as noted above equation 
(\ref{R_perp}), $\alpha\sim 0.5$ in the helium-envelope of GRB progenitor star.

The cross-section for pair annihilation when the average thermal speed of 
$e^\pm$ is $v_\pm$ is $\sigma_T/(v_\pm/c)$. The annihilation time for a 
positron in jet comoving frame, given the density of electrons to be 
$n_\pm'/2$, is therefore 
\be
   t'_{ann} \approx {2\over \sigma_T n'_\pm c}.
\ee
The pair annihilation ceases, and their total number freezes, at a radius where
$t'_{ann} \sim r/c\Gamma(r)$. Thus the freeze-out radius is given by
\be
   R_{freeze} \sim {2\Gamma\over \sigma_T n'_\pm},
    \label{r_freeze1}
\ee
and the pair density at $R_{freeze}$ is
\be
  n'_\pm(R_{freeze}) \approx {2\Gamma(R_{freeze})\over \sigma_T R_{freeze}}.
   \label{n-pair-freeze}
\ee
To determine the pair freeze-out radius we substitute for thermal pair
density, i.e. the following equation
\be
   n'_\pm = {2 (2{\mathrm \pi} k_B m_e T')^{3/2}\over h^3} \exp\left( -m_e c^2/k_B T'
   \right),
\ee
and $r$ dependence of $\Gamma$ and $T'$ into (\ref{r_freeze1}):
\be
\exp\left\{\frac{5.9\times10^9}{T_0}\left(\frac{R}{R_0}\right)
^{\alpha}\right\}\sim \frac{R_0 T_0^{3/2}}{6.2\times 10^8} \left(\frac{R_0}{R}\right)^{5\alpha/2-1}. \label{rfreeze_eqn}
\ee
Let's define 
\be
 C \equiv \frac{5.9\times10^9}{T_0}, \quad{\rm and}\quad
 D \equiv \frac{R_{0}T_0^{3/2}}{6.2\times 10^8}
   \label{coeff_rfreeze},
\ee
and rewrite equation (\ref{rfreeze_eqn}) as
\be
     C\left(\frac{R}{R_0}\right)^\alpha = \log(D)+\left(\frac{5}{2}\alpha-1
    \right)\log\left(\frac{R_0}{R}\right) 
  \label{log_rfr_eqn}.
\ee
which is easier to solve analytically.
Neglecting, at first, the $\log R$-term on the right hand side, we find 
$R=((\log D)/C)^{1/\alpha}$. Substituting that back into (\ref{log_rfr_eqn}), 
the approximate analytical solution for the freeze-out radius is found to be
\be
  \left(\frac{R_{freeze}}{R_0}\right)\sim \frac{1}{C^{1/\alpha}}\left[
  \log(D)+\left(\frac{5}{2}-\frac{1}{\alpha}\right)\left(\log(\log (D))
   -\log(A)\right)\right]^{1/\alpha}
  \label{an_tfr}
\ee
The results for the freeze out radius obtained as exact numerical solution of
equation (\ref{rfreeze_eqn}) are in good agreement with the above approximated expression.
%

The freeze-out radius has a weak dependence on $\alpha$, and for 
$\sigma_0=10^6$ and $T_0=41$ keV, pair density freezes out fairly close to the 
jet launching site. 

The temperature at freeze-out can be calculated using equation 
(\ref{t_erf_nolf}), and it can be shown to be $T'_{freeze} \approx 10$ keV
that is almost independent of various parameters.
The isotropic equivalent luminosity carried by pairs for $r\ge R_{freeze}$ is
\be
  L_\pm^{iso} \approx 4{\mathrm \pi} R_{freeze}^2 m_e c^3 n_\pm' \Gamma^2 \approx
     4{\mathrm \pi} R_{freeze} m_e c^3 \Gamma^3/\sigma_T\approx 
    {4{\mathrm \pi} R_0 m_e c^3\over \sigma_T} \left( {R_{freeze}\over R_0},
    \right)^{1+3\alpha}
\ee
where we made use of equation (\ref{n-pair-freeze}) for pair density at 
$R_{freeze}$.
or
\be
  {L_\pm^{iso}\over L^{iso}} \sim 5\times10^{-16} {R_{0,7}\over L^{iso}_{52}} 
   \left[ 2{L_{52}^{iso}}^{1/4} R_{0,7}^{-1/2} 
    \sigma_{0,6}^{-1/4}\right]^{(3\alpha+1)/\alpha}.
\ee
The jet kinetic luminosity at $r$ is $L/\sigma(r)$, and we see from the above
equation that thermal pairs are insignificant carriers of jet kinetic 
luminosity --- most of the kinetic luminosity is being carried by protons. 
The ratio of $e^\pm$ pair to proton number density above the freeze-out 
radius is given by
\be
  {n_\pm'\over n_p'} \sim 10^{-12} \sigma(R_{freeze}) {L_{52}^{iso}}^{-1}R_{0,7}
   \left[ 2{L_{52}^{iso}}^{1/4} R_{0,7}^{-1/2} \sigma_{0,6}^{-1/4}
    \right]^{(3\alpha+1)/\alpha},
\label{ratio_numdens}
\ee
where $\sigma(R_{freeze})\sim \sigma_0/\Gamma(R_{freeze})$ is the magnetization
 parameter at $r=R_{freeze}$.
Substituting this expression for $\sigma(R_{freeze})$ back into equation 
(\ref{ratio_numdens}) it becomes
\be
  {n_\pm'\over n_p'} \sim 10^{-6} \sigma_{0,6} \left({R_0 \over 
    R_{freeze}}\right)^\alpha {L_{52}^{iso}}^{-1} R_{0,7}
   \left[ 2{L_{52}^{iso}}^{1/4} R_{0,7}^{-1/2} \sigma_{0,6}^{-1/4}
    \right]^{(3\alpha+1)/\alpha},
\label{ratio_numdens_sigma}
\ee

Equation (\ref{ratio_numdens_sigma}) shows that thermal pairs are too small 
in number to affect the propagation of photons into the jet, and hence 
they cannot shield the jet from the severe IC drag.

\subsection{Pair creation due to photon collisions and shielding of jet from 
   IC drag}
\label{pair-create}

The calculation in previous sections ignored the possibility that 
thermal photons IC scattered by the 
jet might have sufficient energy for electron-positron pair creation. These 
IC scattered photons could give rise to an optically thick layer of $e^\pm$ 
on the side wall of the jet that is in contact with the cocoon, and this 
could potentially shield the interior of the jet from IC drag. We investigate 
that possibility in this sub-section.

Let us consider the mean frequency of thermal photons in the cocoon
to be $\nu_c$, and after colliding with an electron in the jet with Lorentz 
factor $\Gamma$ the frequency increases to $\nu_{ic}$ (these frequencies are
in the rest frame of the star). The scattered photon travels within an
angle $\Gamma^{-1}$ of the electron's velocity vector due to relativistic
beaming, and hence its chances of undergoing a second collision with 
another electron is drastically reduced since the photon is moving in 
nearly the same direction as electrons in the jet.

The condition for pair production when averaged over the angle between 
colliding photons is:  $(h\nu_{ic})(h\nu_c) > 2 m_e^2 c^4$, e.g. 
\cite{GouldSchreder:1967}. Since $h\nu_{ic}\sim m_e c^2 \Gamma/2$
(eq. \ref{ic-nu1}) for large angle collisions in the Klein-Nishina 
regime -- i.e. when $h \nu_c \Gamma > m_e c^2$ -- the pair production 
condition becomes $h\nu_c\Gamma \gae 4m_e c^2$. Thus, for a given 
cocoon temperature ($T_c$) at a certain radius where jet Lorentz factor
becomes larger than the following critical value
\begin{equation}
   \Gamma_{crit} \sim {4 m_e c^2\over 3 k_B T_c}
  \label{gam_crit}
\end{equation}
pair production at the interface of the jet and cocoon begins. 
For cocoon temperature of $\sim 10$ keV, $\Gamma_{crit} \sim 70$.
The Lorentz factor of magnetic jets 
increases with radius as $\sim r^{1/2}$, and therefore $\Gamma\sim 10^2$ 
can be attained near the stellar surface where the jet also becomes 
transparent in the transverse direction (in the absence of $e^\pm$) 
for $\sigma_0>10^6$; the Lorentz factor of a thermally driven jet increases more
rapidly with $r$, and it too is likely to develop an $e^\pm$ layer surrounding 
it before reaching the stellar surface.

For the remainder of this section we assume that the condition for pair
production is satisfied, and describe the effect that has on
jet structure and dynamics.

The distance traveled by a high energy IC photon --- which is moving almost 
parallel to the jet axis (within an angle $\Gamma^{-1}$ to be precise) --- 
 before it is turned into $e^\pm$ as a result of collision with a 
thermal photon from the cocoon is
\begin{equation}
  \lambda_\gamma = (\sigma_\pm n_\gamma)^{-1}\sim (2\times 10^2{\rm cm})\,
       t^{1/2} R_{*,11}^{1/2} (L_{50}/\theta_{j,-1})^{-1/4} \eta_{c,1}^{-3/8},
    \label{gam-mean-free-length}
\end{equation}
where $\sigma_\pm$ is the cross-section for $\gamma+\gamma\rightarrow
e^- + e^+$ (the maximum value for $\sigma_\pm$ is $2.5 \times10^{-25}$cm$^2$ at 
photon energy 1.4 times the threshold value given above 
e.g. \cite{Svensson:1982},
\begin{equation}
   n_\gamma(t)\sim {F_c(t)\over h\nu_c c} \sim (1.8\times10^{22} {\rm cm}^{-3})
     \, t^{-1/2} R_{*,11}^{-1/2} (L_{50}/\theta_{j,-1})^{1/4}
     \eta_{c,1}^{3/8},
   \label{n-gam1}
\end{equation}
is the number density of thermal photons at the interface of the cocoon 
and the jet at time $t$ (in seconds) in star rest-frame, $\nu_c=3 k_B T_c$
and $F_c(t)$ are given by equations (\ref{Tc}) and (\ref{Fc}). We see from 
equation (\ref{gam-mean-free-length}) that 
high energy IC photons do not travel very far from their place of creation 
before undergoing pair production.
The newly produced $e^\pm$s have thermal Lorentz factor less than $\sim$2 
in jet comoving frame\footnote{Newly produced pairs are swept up by the
magnetic field in the jet and forced to move with the outflow.} but they
cool rapidly via the synchrotron process on a timescale much smaller than the
dynamical time; magnetic field in the jet comoving frame is
$B'= (4L/\theta_j^2\Gamma^2 r^2c)^{1/2}=(1.2\times10^8 {\rm G}) L_{50}^{1/2}/
  [\theta_{j,-1}\Gamma_2 r_{11}]$, and hence the synchrotron cooling time
in jet comoving frame is $\sim (5\times10^{-8}$s)
  $[\theta_{j,-1}\Gamma_2 r_{11}]^2 L_{50}^{-1}$.

The rate of pair production per electron is approximately 
equal to the number of photons it scatters per unit time, i.e.
\begin{equation}
   {\dot n_\pm\over n_e} \approx {\sigma_T n_\gamma c \over 1 + h\nu_c
   \Gamma/(m_e c^2)} \approx (3\times10^{7} {\rm s}^{-1}) t^{-1/2} \Gamma_2^{-1}
      \eta_{c,1}^{1/2}.
   \label{pair-rate}
\end{equation}
The second equality is obtained for the case where thermal photons from the
cocoon are scattered by electrons in the jet in Klein-Nishina regime, i.e.
$h \nu_c \Gamma/(m_e c^2) = 3 k_B T_c \Gamma/(m_e c^2) \gg1$, and in that
case we find the surprising result that the rate of pair production per
electron depends only on the Lorentz factors of the jet and cocoon\footnote{The
time dependence for pair production rate of $t^{-1/2}$ (eq. \ref{pair-rate}) 
is due entirely to the cocoon thermal flux $F_c$.}.
Once pair production starts it proceeds rapidly since newly produced 
$e^\pm$ also IC scatter thermal photons which have sufficient energy for 
more pair production. The number of pairs produced per proton is expected 
to be of order $m_p\Gamma/(m_e\Gamma_{crit})$; the number is limited by 
pair production rate (which drops to zero when jet LF falls below 
$\Gamma_{crit}$ and IC scattered photons no longer have sufficient energy 
for pair production) and $e^\pm$ annihilation rate.  
Thus, pair production saturates on a timescale 
\begin{equation}
  t_{sat} \sim (3\times10^{-8}{\rm s}) t^{1/2}\,\Gamma_2\, \eta_{c,1}^{-1/2}
     \ln\left[m_p\Gamma/m_e\Gamma_{crit}\right].
\end{equation}
The physical thickness of the pair screen corresponding to optical depth unity,
$\Delta_\pm(\tau_\pm=1)$, is given by
\begin{equation}
     \Delta_\pm(\tau_\pm=1) = {(m_e/m_p) \over \sigma_T n_p 
      (\Gamma/\Gamma_{crit}) } \sim (1.2\times10^5 {\rm cm}) {R_{*,11}^2 
      (\sigma_{0,6}+\xi_{0,6}\theta_{j,-1}^2) \over L_{50}},
   \label{pair-wall-width1}
\end{equation}
where $n_p$ is proton density in the jet at $r=R_*$ which is given by 
equation (\ref{np}), and $\xi_0$ is thermal energy per proton divided
by $m_p c^2$ at the base of the jet.

We note that as long as the Lorentz factor of $e^\pm$ is larger than $\Gamma_{crit}$,
$\gamma$-rays from pair annihilation are converted back to $e^\pm$ within 
a short distance of a few meters (eq. \ref{gam-mean-free-length}) due to
collisions with thermal photons from the cocoon.
Pair production process is terminated when the Lorentz factor of the screen falls
below $\Gamma_{crit}$ and its optical depth becomes so large that X-ray 
photons from the cocoon are prevented from entering the faster
moving interior of the jet where $\Gamma>\Gamma_{crit}$.

Let us consider that the optical depth of $e^\pm$ layer surrounding the jet
at radius $r$ is $\tau_\pm(r)$. The flux of photons from the cocoon that 
crosses this layer is
\begin{equation}
    f_{n_\gamma}(t, \tau_\pm) \approx {F_c(t) \exp(-\tau_\pm)\over 3 k_B T_c}
     \sim (5.5\times10^{38} {\rm cm^{-2}\, s^{-1}}) \exp(-\tau_\pm) 
       {L_{50}^{1/4} \eta_{c,1}^{3/8} \over 
       \theta_{j,-1}^{1/4} R_{*,11}^{1/2}\, t^{1/2} },
\end{equation}
where we made use of equations (\ref{Tc}) and (\ref{Fc}). Pair production
can proceed in the interior to the e$^\pm$ screen as long as the distance
traveled by a IC scattered photon before it collides with a thermal photon
(eq. \ref{gam-mean-free-length}) is smaller than $R_*$, i.e. 
\begin{equation}
  \lambda_\gamma \sim (2\times 10^2{\rm cm})\, e^{\tau_\pm}\,
       t^{1/2}\, R_{*,11}^{1/2}\, (L_{50}/\theta_{j,-1})^{-1/4}
     \eta_{c,1}^{-3/8}<R_*.
    \label{gam-mean-free-length-new}
\end{equation}
Moreover, the time it takes for a photon 
from the cocoon to cross this layer should also be less than $R_*/c$.
Thus, we find that the optical depth of the screen is
\begin{equation}
   \tau_\pm \sim 20 + \log\left[ t^{-1/2}\,R_{*,11}^{-1/2} 
    (L_{50}/\theta_{j,-1})^{1/4} \eta_{c,1}^{3/8} \right],
\end{equation}
and the physical thickness of the pair screen is $\max\{20 \Delta_\pm,
 R_*/\Gamma\}$, where $\Delta_\pm$ is given by equation 
(\ref{pair-wall-width1}); pair screen thickness is $R_*/\Gamma$ when IC 
photons --- moving at an angle $\Gamma^{-1}$ with respect to the jet axis ---
travel a distance $R_*$ before colliding with thermal photons from the cocoon.

\begin{figure}
\centering
\begin{tikzpicture}[scale=1]
\shadedraw[left color=white, right color=yellow!15!white,draw=none](0,0) --node[below, sloped, very near end,font=\bfseries, color=gray!70!white]{star envelope} +(30:9) arc(30:-30:16.5 and 9) -- cycle;
\shadedraw[left color=cyan!30!white, right color=blue!40!cyan,draw=none] plot [smooth cycle] coordinates {(0,0) (8,2.5) (10.2,0) (8,-2.5)}; 
\shadedraw[color=red!20!white,left color=red!20!white, right color=red!80!magenta, style=very thin, draw=none]
plot [smooth cycle] coordinates {(0,0) (8,1.55) (10,0) (8,-1.55)};
\shadedraw[color=red!20!white,left color=red!20!white, right color=red!80!magenta, style=very thin, draw=none]
plot [smooth cycle] coordinates {(2,0) (8.5,1.9) (10.05,0) (8.5,-1.9)};
\shadedraw[color=orange!20!white,left color=orange!20!white, right color=orange!50!white, style=very thin, draw=none]
plot [smooth cycle] coordinates {(0,0) (8,1.4) (10,0) (8,-1.4)};
\shadedraw[left color=yellow!20!white, right color=yellow!50!white, style=very thin, draw=none] plot [smooth cycle] coordinates {(0,0) (8,.8) (10,0) (8,-.8)}; 
\begin{scope}[color=black,style=very thick]
\tikzstyle{every node} = [scale=.25,fill=red!70!, circle,text=black,font=\large,font=\bfseries, style=very thick]
\node at (1.8,0.4) {$+$};
\node at (2.5,0.5) {$+$};
\node at (3,0.55) {$+$};
\node at (3.8,0.65) {$+$};
\node at (4,.75) {$+$};
\node at (4.6,.7) {$+$};
\node at (5.2,.9) {$+$};
\node at (5.5,.85) {$+$};
\node at (6.,0.9) {$+$};
\node at (6.5,1.15) {$+$};
\node at (6.7,.9) {$+$};
\node at (7.1,1.05) {$+$};
\node at (7.6,1.0) {$+$};
\node at (8,1.15) {$+$};
\node at (8.3,1) {$+$};
\node at (8.8,1.1) {$+$};
\node at (8.9,.9) {$+$};
\node at (9.25,.85) {$+$};
\node at (9.65,.6) {$+$};
\node at (1.8,-0.4) {$+$};
\node at (2.5,-0.5) {$+$};
\node at (3,-0.55) {$+$};
\node at (3.8,-0.65) {$+$};
\node at (4,-.75) {$+$};
\node at (4.6,-.7) {$+$};
\node at (5.2,-.9) {$+$};
\node at (5.5,-1.0) {$+$};
\node at (6.,-0.9) {$+$};
\node at (6.5,-1.2) {$+$};
\node at (6.7,-.9) {$+$};
\node at (7,-1.2) {$+$};
\node at (7.1,-1.05) {$+$};
\node at (7.6,-1) {$+$};
\node at (8,-1.2) {$+$};
\node at (8.3,-1) {$+$};
\node at (8.8,-1.1) {$+$};
\node at (8.9,-.9) {$+$};
\node at (9.25,-.85) {$+$};
\node at (9.6,-.6) {$+$};
\tikzstyle{every node} = [scale=.25,fill=blue!30!white, circle,text=black,font=\large,font=\bfseries, style=very thick]
\node at (1.33,0.35) {$-$};
\node at (2.3,0.45) {$-$};
\node at (2.85,0.5) {$-$};
\node at (3.4,.6) {$-$};
\node at (4.3,.7) {$-$};
\node at (4.7,.9) {$-$};
\node at (5.7,1.1) {$-$};
\node at (5.8,0.8) {$-$};
\node at (6.,1.15) {$-$};
\node at (6.4,0.9) {$-$};
\node at (8,1.) {$-$};
\node at (7.4,1.1) {$-$};
\node at (7.8,1.25) {$-$};
\node at (8.4,1.1) {$-$};
\node at (8.6,1.) {$-$};
\node at (9.05,1.) {$-$};
\node at (9.4,.7) {$-$};
\node at (9.75,.45) {$-$};
\node at (1.33,-0.35) {$-$};
\node at (2.3,-0.45) {$-$};
\node at (2.85,-0.5) {$-$};
\node at (3.4,-.6) {$-$};
\node at (4.3,-.7) {$-$};
\node at (4.7,-.9) {$-$};
\node at (5.7,-1.1) {$-$};
\node at (5.8,-0.8) {$-$};
\node at (6.1,-1.1) {$-$};
\node at (6.5,-1.0) {$-$};
\node at (6.7,-1.1) {$-$};
\node at (8.2,-1.2) {$-$};
\node at (7.4,-.9) {$-$};
\node at (7.6,-1.2) {$-$};
\node at (7.8,-.9) {$-$};
\node at (8.4,-1.15) {$-$};
\node at (8.6,-1.) {$-$};
\node at (9.05,-1.) {$-$};
\node at (9.4,-.7) {$-$};
\node at (9.75,-.45) {$-$};
\end{scope}
\draw (10.5,3) -- (10.2,3) --node[very near start, above=4pt, right=8pt,font=\bfseries] {cocoon} (9.1,1.5);
\draw (10.5,2.5) -- (10.2,2.5) --node[very near start, above=4pt, right=8pt,font=\bfseries] {IC-dragged layer} (9.3,1.1);
\draw (10.5,2) -- (10.2,2) --node[very near start, above=4pt, right=8pt,font=\bfseries] {pairs-layer} (9.5,.7);
\draw (10.5,1.5) -- (10.2,1.5) --node[very near start, above=4pt, right=8pt,font=\bfseries] {jet core} (9.7,.3);
\draw[black, dotted, very thick,->] (0,0)-- (12,0);
\begin{scope}[xshift=-5cm,yshift=1cm]
\pgfsetfillopacity{.8}
\draw[fill=white,style=very thin] (2,-.2)--(5.5,-.2)--(5.5,2)--(2,2)--(2,-.2);
\draw[->,decorate,decoration={snake,amplitude=.5mm,segment length=2.5mm}]
 {(2.7,1.1) -- node[->,very near start, above=-1pt]{$\gamma_{c}$} (3,0.3)};
\node[scale=.3,fill=yellow, circle,text=black,font=\large,font=\bfseries, style=very thick,color=black] at (3.1,.2) {$-$};
\node[scale=.29,fill=yellow, circle,text=black,font=\large,font=\bfseries, style=very thick] at (3.1,.2) {$-$};
\draw[->,decorate,decoration={snake,amplitude=.1mm,segment length=1.5mm}]
 {(3.2,.2) -- node[midway, above]{$\gamma_{ic}$} (4.5,0.2)};
\draw[->,decorate,decoration={snake,amplitude=.5mm,segment length=2.5mm}]
 {(4,1.4) -- node[very near start, above=1pt]{$\gamma_{c}$} (4.5,0.25)};
\draw[->](4.55,0.25) --(4.88,0.65);
\draw[->](4.55,0.2) --(5.15,0.7);
\node[scale=.31,fill=black, color=black,circle] at (4.95,.75) {$+$};
\node[scale=.31,fill=black, color=black,circle] at (5.2,0.8) {$-$};
\node[scale=.3,fill=red!80!, circle,text=black,font=\large,font=\bfseries, style=very thick] at (4.95,.75) {$+$};
\node[scale=.3,fill=blue!60!white, circle,text=black,font=\large,font=\bfseries, style=very thick] at (5.2,0.8) {$-$};
\draw[->,thick] (5.6,.9) -- (9.5,-.2);
\end{scope}
\end{tikzpicture}
\caption{Schematic view of the layered jet structure while it has been braked by the Inverse Compton interaction with the hot photons from the cocoon. If the magnetization parameter $\sigma$ is larger than $10^7$, the jet build up a self-shielding $e^+/e^-$-pair screen created by photons from the cocoon and IC-scattered photons.}
\label{pair-screen}
\end{figure}
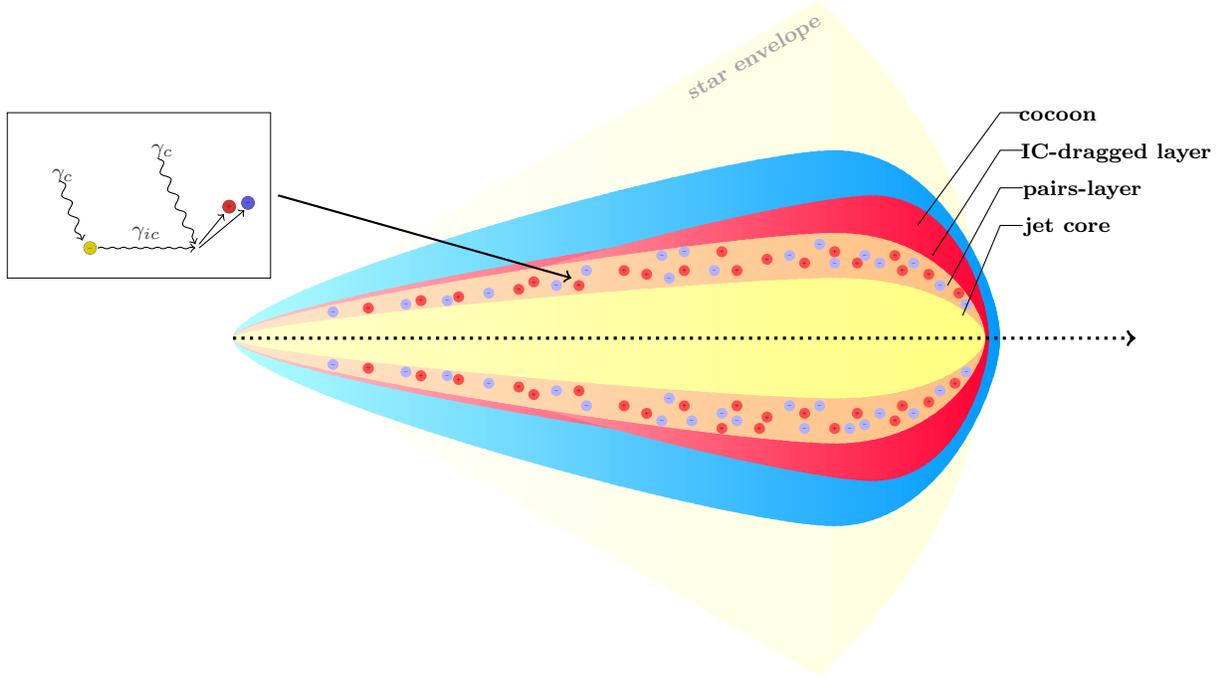

Electrons and positrons in the pair screen continue to scatter X-ray photons
from the cocoon and the resulting drag slows down the jet below 
$\Gamma_{crit}$ (fig. \ref{pair-screen}). 
The time it takes for $\Gamma$ to fall below $\Gamma_{crit}$ is of order
10$^{-4}$s $\exp(\tau_\perp)/\zeta_\pm$ (eq. \ref{t-ic-kn}).
When the Lorentz factor of pairs falls below $\Gamma_{crit}$ (eq. \ref{gam_crit}),
$\gamma$-rays produced in pair annihilation no longer have
sufficient energy for pair creation, and at that time the pair screen begins
to evaporate.

The time (measured in the rest frame of the star) for a positron to run 
into an electron and annihilate is
\begin{equation}
   t_{e^\pm,an}(r) = {\Gamma^2\over \sigma_{e^\pm -> \gamma\gamma} n_\pm v_\pm}
    = {\Gamma^2 \over \sigma_T n_\pm c} \sim (5\times10^2 {\rm s}) {\Gamma_2^2
   r_{11}^2 (\sigma_{0,6} + \xi_{0,6})\over \zeta_\pm L_{52}^{iso} },
   \label{t-annihilate}
\end{equation}
where $\sigma_{e^\pm -> \gamma\gamma} = \sigma_T/(v_\pm/c)$ is annihilation 
cross-section, $v_\pm$ is the average relative speed between electrons 
and positrons in jet comoving frame, and $n_\pm \sim \zeta_\pm n_p$ is the
pair density; for $\zeta_\pm \sim (m_p/2m_e) (\Gamma/\Gamma_{crit})$, i.e.  
the maximum possible number of pairs per protons when $\Gamma >\Gamma_{crit}$, 
the annihilation time is $\sim 0.3$ s.  The ratio of annihilation and dynamical
times at radius $r$ is
\be
   {t_{e^\pm,an}(r)\over t_{dyn}} \sim 1.5\times10^2 {\Gamma_2^2
   r_{11} (\sigma_{0,6} + \xi_{0,6})\over \zeta_\pm L_{52}^{iso} },
  \label{t-annihilate-ratio}
\ee
which is smaller than 1 for $r\lae 10^9$cm even for $\zeta_\pm=1$, and
that means that any pairs produced at the base of the jet or at any radius 
smaller than $10^9$cm cannot survive as the jet moves to larger radii. Hence,
only an ongoing process of pair formation can support $\zeta_\pm > 1$.

The Lorentz factor of pair screen continues to decrease due to IC drag and that causes
the annihilation time ---  which scales as $\Gamma^2$ (eq. 
\ref{t-annihilate})\footnote{For a time-independent relativitic outflow, 
particle density in star rest frame does not vary as we follow the 
flow-steamlines even though the Lorentz factor might increase or decrease. 
This is due to the conservation of particle flux and the fact that the speed 
is a constant $c$ for a relativistic system. The particle density along a 
flow-line, of course, varies as $\Gamma^{-1}$ in the comoving frame of the 
outflow.}
--- to decrease rapidly. Since the
timescale for the creation of pair screen is smaller than 
the IC drag time which is much smaller than the pair annihilation time,
soon after the pair screen forms its Lorentz factor decreases rapidly to order
unity (in 1 $\mu$s or less -- eq. \ref{t-ic-ts}), pairs annihilate
and screen evaporates on timescale of order 1 ms (eq. \ref{t-annihilate}).
Once the optical depth of the screen decreases, photons from the cocoon 
pass through it, and the process of formation
of $e^\pm$ and IC drag progresses deeper inside the jet. At any given
radius this process is only terminated at a distance from jet axis
where the Thompson optical depth out to the jet-cocoon boundary, for 
$\zeta_\pm \sim 1$, is of order 10. Hence, as we get closer to the
stellar surface we find an increasingly larger fraction of the jet to have
gone through the process of forming pair screen, e$^\pm$ annihilation,
and IC drag. If the jet with $\zeta_\pm=1$ becomes transparent in the 
transverse direction near the stellar surface -- as it in fact does 
for $\sigma_0\gae 10^6$ according to eq. \ref{r_ph} -- then self-generated 
pair screen is too short lived to protect it from IC drag.
It should be noted that the transient nature of the pair screen in fact
speeds up the process of jet IC drag because the drag time is proportional
to $\zeta_\pm^{-1}$ (see eqs. \ref{t-ic-kn}-\ref{t-ic-ts}).

Although very high-$\sigma$ jets ($\sigma_0\gae 10^6$) are unlikely to 
survive their passage through the GRB progenitor star and cocoon, they
do not disappear without leaving an observational signature. 
Annihilation of pairs in the screen when the jet Lorentz factor decreases from 
$\Gamma_{crit}$ to of order unity produces photons of energy 
between 0.5 MeV and $\sim m_e c^2 \Gamma_{crit} \sim 50$ MeV
which can escape the jet in the longitudinal direction to arrive
at the observer. The jet is slowed down by pair creation and IC drag, and these processes
are about equally important for decelerating the jet. Hence, the observer-frame luminosity carried by pairs is of order the jet luminosity ($L_j$).
The total energy carried by the photons resulting from pair annihilation
is of the order of the energy of the pairs, and therefore, the total luminosity of pair annihilation photons, $L_\gamma$, is of the order $L_j$.
The duration of the annihilation pulse we expect to be of order the activity time of the central engine, which for long GRBs is typically around 5-100 s.

The picture that emerges is that the outer layers of GRB jets are slowed
down due to IC drag. However, inner regions continue to accelerate with $r$, 
and at some radius their Lorentz factor exceeds $\Gamma_{crit}$. At that point a very 
rapid generation of $e^\pm$ ensues provided that the optical depth of the
slower moving layer outside of this region is less than about 20.
The IC drag slows down the pair screen rather quickly and then $e^\pm$
annihilate on a relatively short time scale of $\sim $1 ms, and the formation
of a new pair screen moves closer to the jet axis. This process continues until 
the jet becomes opaque in the transverse direction due to just the
electrons associated with protons, i.e. for $\zeta_\pm=1$. Jets 
of initial magnetization
($\sigma_0$) smaller than 10$^6$ are sufficiently opaque in transverse
direction even when they rise above cocoon surface that they are essentially 
protected from IC drag. However, photons from the cocoon can penetrate 
a jet all the way to its axis when $\sigma_0\gae 10^6$, and the strong IC drag 
then slows down the outflow to sub-relativistic speed. It turns out that 
high-$\sigma$ jets cannot escape this fate in spite of the e$^\pm$ pair
screen they create because these screens are rather short lived.

\section{Conclusions}

Relativistic jets in GRBs are surrounded by a hot cocoon of plasma that 
was created during the initial passage of the jet through the star when it
shock heated the gas along its path and pushed it sideways 
to clear a cavity through the polar region of the
GRB progenitor star. Thermal photons from
this cocoon are scattered by electrons in the jet and that provides
a strong drag force on the jet. Jets of initial magnetization parameter
($\sigma_0$) smaller than about 10$^6$ are highly opaque in the transverse
direction while traveling inside the GRB progenitor star, and thus they have 
a core region that is protected from this IC drag. The outer layers
of this jet (about 20 Thompson optical depth thick), however, suffer IC drag 
and are slowed down considerably. 

Jets with $\sigma_0 \gae 10^6$ are transparent to photons from the cocoon
near the stellar surface, and they are slowed down to sub-relativistic 
speeds due to IC drag. This is in spite of the fact that an optically thick 
layer of electrons and positrons forms at the interface of the cocoon and 
jet and that tries to protect the core of the jet from IC drag 
(fig. \ref{pair-screen}); these pairs 
are formed by the collisions of thermal photons from the cocoon with high 
energy photons that are produced when cocoon photons are IC scattered by 
the jet.  However, the problem is that the pair screen itself slows down 
rapidly due to IC drag, and that causes pairs to annihilate and the 
$e^\pm$-shield to evaporate rather quickly. Pair production then moves 
closer toward the jet axis, and the story is repeated until the
entire jet is slowed down by the IC drag.

The process of pair-screen formation and annihilation associated with a
high-$\sigma$ jet has observational consequences. 
For jets with $\sigma_0 \gae 10^6$ we should see a pulse of 
high energy photons of energy between $\sim$ 1 MeV and 
$\sim m_e c^2 \Gamma_{crit} \sim 50$ MeV with luminosity of order
the Poynting jet luminosity. The duration of this pulse should be of 
order the central engine activity time.

Even Poynting jets of $\sigma_0\lae 10^6$ -- which are highly opaque 
in the transverse direction --- might suffer effects of IC drag indirectly. 
As the outer layers of these jets are slowed down by IC scatterings
the resulting shear instabilities might slow down the inner regions as well.
Moreover, if magnetic field lines thread the outer and the inner regions
of the jet then slowing down of the outer part of the jet would get
communicated to the inner region, and that could effect the entire jet.
The details of this would depend on the magnetic field configuration
and that is something that needs to be looked into. 

If a highly magnetized jet manages to escape IC drag while traveling 
inside the GRB progenitor star -- for instance if it is encapsulated
inside a highly opaque baryonic outflow to shield it from X-rays from
the cocoon  -- it would be subjected to rapid dissipation due to charge 
starvation before reaching the deceleration radius (appendix B).

\section*{\centering Acknowledgments}
We thank Jonathan Granot for useful discussions.
This work was funded in part by NSF grant ast-0909110.

\vfill\eject
\appendix
\numberwithin{equation}{section}
\bigskip
\section{Radiation from cocoon surrounding relativistic jet}
\medskip

We describe in this appendix a simplified, analytical, treatment of 
cocoon dynamics and radiation. 
This follows the work of \cite{RRCR02} and \cite{Matzner:2003}
except for one thing
and that is that we do not assume that the jet is conical in shape while
inside the star. Numerous investigations of relativistic jets, e.g.
\cite{Lazzati:2005}, \cite{Morsony:2007}, \cite{mizuta:2009},
\cite{mizuta:2013}, \cite{Bromberg:2011,Bromberg:2014} have shown that the 
cocoon created by the jet can be very effective in collimating it, and hence
we take the jet opening angle to be a function of distance from the center.

The energy in the cocoon ($E_c$) is of order the energy carried by the jet 
while it makes its way through the polar region of the GRB progenitor star,
i.e. 
\be
  E_c\sim L R_*/v_h,
  \label{E_c1}
\ee
where $v_h$ is the average speed at which the jet head moves through the star. 
The jet head speed can be calculated from the conservation of momentum flux 
in the radial direction of the unshocked stellar gas as viewed from the 
rest frame of the jet head:
\be
 \rho_j c^2 (\Gamma_j^2/4\Gamma_h^2) \approx \rho_a \Gamma_h^2 v_h^2,
\ee
where $\rho_j$ and $\rho_a$ are densities of the unshocked jet and the 
stellar envelope respectively, and $\Gamma_j$ and $\Gamma_h$ are the Lorentz
factors of the unshocked jet and the jet-head with respect to the unshocked star
(the Lorentz factor of the unshocked jet with respect to jet-head is $\Gamma_j/2\Gamma_h$).
Considering that the jet luminosity at the stellar surface can be written as
\be
   L = {\mathrm \pi} \theta_j^2 R_*^2 \rho_j \Gamma_j^2 c^3,
  \label{L_j}
\ee
and the mass of the swept up gas by the jet is
\be
   m_c \sim {\mathrm \pi} \theta_j^2 \rho_a R_*^3,
  \label{m_c}
\ee
we obtain
\be
  2 \Gamma_h^2 v_h \sim \left[ {R_* L\over c m_c}\right]^{1/2},
  \label{vh}
\ee
where $\theta_j$ is jet opening angle which is a function of distance from
the center due to collimation provided by the cocoon.
We can simplify this expression further by substituting for $L$ using
equation (\ref{E_c1})
\be 
  4 \Gamma_h^4 v_h \sim c \eta_c
  \label{vh1}
\ee
where
\be
   \eta_c \equiv {E_c\over m_c c^2}
  \label{eta_c}
\ee
is the terminal Lorentz factor of the cocoon plasma (provided that
$\eta_c\ge1$) after it escapes through the stellar surface and its 
thermal energy is converted to bulk kinetic energy.\\
Therefore, the jet head speed is sub-relativistic when $\eta_c< 4$ and
is given by
\be
   v_h \sim c\eta_c/4.
   \label{vh2a}
\ee
For $\eta_c>4$, the jet head speed is relativistic and its Lorentz factor
is given by
\be
   \Gamma_h \sim (\eta_c/4)^{1/4}.
   \label{vh2b}
\ee
The expansion speed of the cocoon in the
direction perpendicular to its surface, $v_c$, is determined by equating
the ram pressure with the thermal pressure inside the cocoon ($p_c$),
e.g. \cite{Matzner:2003}, 
\be
  v_c = (p_c/\rho_a)^{1/2}.
  \label{vc1}
\ee
The average thermal pressure inside the cocoon is approximately
\be
  p_c \sim {E_c\over 3 V_{c}}.
  \label{p_c}
\ee
where 
\be
 V_{c}\sim {\mathrm \pi} t^3 v_h v_c^2/3 \sim R_*^3 (v_c/v_h)^2
 \label{cocoon-volume}
\ee
 is the volume of the cocoon
at the time it emerges at the stellar surface. Combining equations (\ref{vc1})
and (\ref{p_c}) we find 
\be
  v_c^4\sim {E_c v_h^2 \over 3\rho_a R_*^3} \sim \theta_j^2 \eta_c v_h^2 c^2/3.
  \label{vc2}
\ee
Substituting for $v_h$ from equations (\ref{vh2a}) and (\ref{vh2b}) we find
\begin{equation}
 {v_c\over c} \sim \left\{
 \begin{array}{ll}
   (\theta_j^2 \eta_c^3/48)^{1/4}  & {\rm for}\; \eta_c<4 \\
   (\theta_j^2 \eta_c/3)^{1/4}     & {\rm for}\; 4<\eta_c<\theta_j^{-2}
 \end{array}
\right.
   \label{vc3}
\end{equation}
The thermal pressure of the cocoon can be obtained using equations
(\ref{vc1}) and (\ref{vc3}) and is given by
\be
  p_c \sim {L \over \theta_j R_*^2 c (3\eta_c)^{1/2} },
  \label{pc}
\ee
and its temperature is
\be
   k_B T_c = k_B (3 p_c/\sigma_a)^{1/4} \sim (24 {\rm keV}) 
   { L_{50}^{1/4} \over \theta_{j,-1}^{1/4}\,R_{*,11}^{1/2} \eta_{c,1}^{1/8} },
   \label{Tc}
\ee
where $\sigma_a$ is the radiation constant. We note that the cocoon 
temperature has a weak dependence on jet luminosity and angular size, and
so it is unlikely to be larger than $\sim 30$ keV.\\
The number density of thermal $e^\pm$ pairs at temperature $T_c$ is given by
\be
   n_\pm = {2 (2{\mathrm \pi} k_B m_e T_c)^{3/2} \over h^3} \exp\left(-m_e c^2/k_B T_c
   \right).
\ee
Therefore, for $k_B T_c = 20$ keV, $n_\pm = 1.1\times10^{17}$cm$^{-3}$,
and for 30 keV cocoon temperature $n_\pm=1.1\times10^{21}$cm$^{-3}$.
We next calculate the number of electrons associated with protons in the 
cocoon, and show that these exceed $e^\pm$ as long as $k_B T < 30$ keV.\\
The average number density of electrons associated with baryons in the cocoon is
\be
  n_{e,c} \sim {m_c\over m_p V_{c}} \sim {3 p_c \over m_p c^2 \eta_c} 
   \sim (1.5\times10^{21}{\rm cm}^{-3}) { L_{50}\over \theta_{j,-1} 
  R_{*,11}^2 \eta_{c,1}^{3/2} }.
\ee
The electron number density near the stellar surface, however, is smaller than
the average density given above. The density at the stellar photosphere is 
$\sim 1/(\sigma_T H_\rho)$, where 
\be
  H_\rho = C_s^2/g  \sim (10^8{\rm cm}) T_{*,5} R_{*,11}^2 M_{*,1}^{-1},
\ee 
is the density scale height, $C_s$ is sound
speed, $T_*$ is photospheric temperature, and $M_{*,1}$ is stellar mass
in units of 10 $M_\odot$. Thus, the electron density at the photosphere
is of order $1.5\times10^{16}$ cm$^{-3}$, and it increases with depth as
$(z/z_*)^{1/(\gamma-1)}$; where $\gamma\sim1.5$ is the effective polytropic 
index that describes stratification near the stellar surface, and 
\be
  z_*=H_\rho/(\gamma-1).
\ee
Therefore, pair density in cocoon is larger than proton density 
at the photosphere as long as $k_B T > 15$ keV, but at a depth of more than
a few scale height below the photosphere the proton density exceeds $n_\pm$.
So the electron density in the cocoon as a function of radius can be written
as
\be
  n_{e,c} \sim n_\pm + \min\left\{ \left({1\over\sigma_T H_\rho}\right)\left[1 + {(R_*-r)\over z_*}\right]^{1/(\gamma-1)}
  , \; (1.5\times10^{21}{\rm cm}^{-3}) { L_{50} \over 
  \theta_{j,-1} R_{*,11}^2 \eta_{c,1}^{3/2} } \right\}. \label{el-den}
\ee
The time in between scattering for a photon (photon mean free time) in the 
cocoon is given by
\be
   t_{fs} \sim {1\over \sigma_T~ c~ n_{e,c}}.
   \label{tfs}
\ee
The thermal flux at the interface of the cocoon and the jet is dictated 
by diffusion of photons in cocoon, and is given by:
\be
   F_c(t) = \sigma_B T_c^4 [t_{fs}/(t+t_{fs})]^{1/2}.
   \label{Fc}
\ee
This expression is valid as long as $t$ is less than the cocoon expansion time 
$\sim R_*/v_h$.  The flux at an optical depth $\tau$ inside the jet is 
$f_c(t) \exp(-\tau)$.

\vfill\eject
\medskip
\section{Charge starvation of a Poynting jet}
\medskip

Let us consider a Poynting jet of isotropic equivalent luminosity $L^{iso}$ 
and magnetization parameter at its base of $\sigma_0$. The magnetic field
in the jet is assumed to change direction on a length scale of $\ell_B$
(in star rest frame) which corresponds to $\ell_B' = \ell_B \Gamma$ in the
jet comoving frame. The current required for supporting this non-zero 
curl is 
\be
   j' \sim B' c/(4{\mathrm \pi} \ell_B'),
\ee
where 
\be
   B' = {1\over \Gamma} \left[ {L^{iso}\over c r^2}\right]^{1/2}
\ee
is magnetic field in jet comoving frame.

The electron density in jet comoving frame for a Poynting jet of high 
magnetization parameter is obtained using equation (\ref{np-prime}) and 
is given by
\be
  n_e'(r) \approx \frac{\zeta_\pm L^{iso}}{4{\mathrm \pi} r^2 m_p c^3 \sigma_0 \Gamma},
\ee
where $\zeta_\pm$ is the number of $e^\pm$ per proton which we know from 
the discussion in section \ref{ep-p} should be of order unity for $r\gg
R_*$.

The current required to support the jet magnetic field must be
smaller than the maximum current that can be carried by charged particles 
in the jet, i.e. $j' < q\,n_e' c$. It follows from this requirement
that beyond a certain radius, $R_{cs}$, the jet becomes charge starved, 
i.e. it does not have sufficient number of electrons to carry the required 
current. Using the above equations we find this radius to be
\be
   R_{cs} \sim {q \ell_B \Gamma \sqrt{L^{iso}}\over m_p \sigma_0 c^{5/2}}
    \sim (1.5\times10^{17}{\rm cm}) \sqrt{L^{iso}_{52}} {\ell_{B,7} 
    \Gamma_2\over \sigma_{0,6}}. 
\ee
This should be compared with the deceleration radius, $R_d$, where the energy 
of the medium swept-up and shock heated by the jet is approximately
half the total energy of the explosion:
\be
  R_d \sim \left( {3 E^{iso} \over 4{\mathrm \pi} m_p c^2 n_0 \Gamma^2} \right)^{1/3}
   \sim (1.2\times10^{17}{\rm cm}) \left[ {E^{iso}_{53} \over n_0 \Gamma_2^2}
    \right]^{1/3},
\ee
where $E^{iso}$ is the isotropic equivalent of total energy carried by the jet,
and $n_0$ is the mean number density of protons in the circum-stellar medium 
of the GRB. Thus, a high magnetization jet that stops accelerating when it
attains a Lorentz factor of $\sim\sigma_0^{1/3}$ will become charge starved near the
deceleration radius and its magnetic field will dissipate rapidly.

We note that if the jet were to start spreading in the radial direction at 
$r < R_{cs}$ then it would never become charge starved since in that case
the required current ($j'$) and the charge density ($n_e'$) both decline
with radius as $r^{-2}$.

\vfill\eject

\bibliographystyle{mn2e}
\bibliography{myGRBbiblio}

\end{document}